\documentclass[fleqn,usenatbib]{mnras}

\usepackage{newtxtext,newtxmath}

\usepackage[T1]{fontenc}

\DeclareRobustCommand{\VAN}[3]{#2}
\let\VANthebibliography\thebibliography
\def\thebibliography{\DeclareRobustCommand{\VAN}[3]{##3}\VANthebibliography}

\usepackage{graphicx}	
\usepackage{amsmath}	

\title[POSS I vanishing objects]{Discovering vanishing objects in POSS I red images using the Virtual Observatory}

\author[E. Solano et al.]{
Enrique Solano,$^{1,2}$\thanks{E-mail: esm@cab.inta-csic.es}
B. Villarroel,$^{3,4}$
C. Rodrigo$^{1,2}$
\\
$^{1}$Centro de Astrobiolog\'{\i}a (CAB), CSIC-INTA, Camino Bajo del Castillo s/n, E-28692, Villanueva de la Ca\~{n}ada, Madrid, Spain\\
$^{2}$Spanish Virtual Observatory\\
$^{3}$Nordita, KTH Royal Institute of Technology and Stockholm University, Roslagstullsbacken 23, SE-106 91 Stockholm, Sweden\\
$^{4}$ Instituto de Astrof\'isica de Canarias, Avda V\'ia L\'actea S/N, La Laguna, E-38205, Tenerife, Spain\\
}

\date{Accepted 2022 May 31. Received 2022 May 03; in original form 2022 March 08}

\pubyear{2022}

\begin{document}
\label{firstpage}
\pagerange{\pageref{firstpage}--\pageref{lastpage}}
\maketitle

\begin{abstract}
In this paper we report a search for vanishing sources in POSS I red images using Virtual Observatory archives, tools and services. The search, conducted in the framework of the VASCO project, aims at finding POSS I (red) sources not present in recent catalogues like Pan-STARRS DR2 (limiting magnitude $r$=21.4) or \emph{Gaia} EDR3 (limiting magnitude $G$=21). We found 298\,165 sources visible only in POSS I plates, out of which 288\,770 had a crossmatch within 5\,arcsec in other archives (mainly in the infrared), 189 were classified as asteroids, 35 as variable objects, 3\,592 as artefacts from the comparison to a second digitization (Supercosmos), and 180 as high proper motion objects without information on proper motion in {\emph Gaia} EDR3. The remaining unidentified transients (5\,399) as well as the 172\,163 sources not detected in the optical but identified in the infrared regime are available from a Virtual Observatory compliant archive and can be of interest in searches for strong M-dwarf flares, high-redshift supernovae, asteroids, or other categories of unidentified red transients. No point sources were detected by both POSS-I and POSS-II before vanishing, setting the rate of failed supernovae in the Milky Way during 70 years to less than one in one billion. 
\end{abstract}




\section{Introduction}
Transients can be defined as astrophysical phenomena whose duration is significantly lower than the typical timescale of the stellar and galactic evolution (from seconds to years in contrast to millions or billions of years). Supernovae, novae, gamma-ray bursts,..., are some examples of transient events. 

While most of the modern surveys have developed robust and well tested facilities to discover transients (ASAS \citep{Poj97}, OGLE \citep{Udalski15}, ZTF \citep{Bellm19} or \emph{Gaia} \citep{Gaia21}, to name a few), the same level of attention is not paid to vanishing events, that is, objects detected in old surveys but that are not identified in more recent ones. In this context, vanishing refers both to known types of objects faded below the detection limits (e.g. large amplitude variables) as well to unknown physical phenomena non predicted before. Objects that do not appear in the same position because they have moved in the sky (e.g. solar system objects) do not belong to this category. 

VASCO\footnote{\url{https://vasconsite.wordpress.com/}} (Vanishing and Appearing Sources during a Century of Observations) is a project to search for vanishing and appearing sources using existing survey data. Examples of exceptional astrophysical transients found with VASCO are given in \citet{Villarroel20}. VASCO also runs a citizen science project that investigates 150 000 candidates visually \citep{Villarroel20b}. This paper, conceived in the frame of the VASCO project, aims at performing an automated search for vanishing object using the digitized plates of the First Palomar Sky Survey (POSS I) and the Pan-STARRS (DR2) and \emph{Gaia} EDR3 catalogues taking advantage of Virtual Observatory (VO) tools and services\footnote{\url{https://www.ivoa.net/}}. 

POSS I was conducted between 1949--1956 (99 per cent of the plates) and 1956--1958 (1 per cent  of the plates) using the 48-inch Oschin Schmidt telescope at Mount Palomar in southern California \citep{Minkowski63}. The survey covers the entire sky north of  -45\,deg declination and was carried out using photographic plates, later converted into a digital format. In order to obtain colour information, each region of the sky was photographed twice, once using a blue sensitive Kodak 103a-O plate, and once with a red sensitive Kodak 103a-E plate, peaking at $\sim$ 4\,100 and $\sim$ 6\,400\,\AA, respectively\footnote{\url{https://authors.library.caltech.edu/31250/1/Palomar_Observatory_Sky_Atlas.pdf}}. The limiting photographic magnitudes of the blue and red plates are 21.1 and 20.0 mag, respectively.


In this paper we made use of the digitization of the POSS I red plates available from the Digitized Sky Survey from ESO\footnote{\url{http://archive.eso.org/dss/dss/}} and produced at the Space Telescope Science Institute through its Guide Star Survey group\footnote{\url{http://gsss.stsci.edu/Catalogs/Catalogs.htm}}. 

The Panoramic Survey Telescope and Rapid Response System (Pan-STARRS\footnote{\url{https://panstarrs.stsci.edu/}}) is a system for wide-field astronomical imaging developed and operated by the Institute for Astronomy at the University of Hawaii. Pan-STARRS1 (PS1) is the first part of Pan-STARRS.  PS1, located at Haleakala Observatory, started operation in 2010 and have surveyed the sky in five bands ($g$,$r$,$i$,$z$,$y$) at declinations higher than -30\,deg using a 1.8 meter telescope and a 1.4 Gigapixel camera. 
The PS1 Second Data Release (DR2) survey contains almost 2 billion objects and was released on January 28, 2019. PS1 (DR2) was queried using the corresponding Virtual Observatory ConeSearch service\footnote{\url{http://gsss.stsci.edu/webservices/vo/CatalogSearch.aspx?&CAT=PS1V3OBJECTS&RA=&DEC=&SR=}}.

\emph{Gaia} is a European space mission providing astrometry, photometry and spectroscopy of more than 1\,500 million stars in the Milky Way. Also data for significant samples of extragalactic and Solar system objects are made available. Photometric information is provided in thre bands: $G$, $G_{BP}$ and $G_{RP}$ with a limiting magnitude of $G$ $\sim$ 21 mag\footnote{\url{https://www.cosmos.esa.int/web/gaia/earlydr3}}.  \emph{Gaia} EDR3 \citep{Gaia21} is based on data collected between 25 July 2014 and 28 May 2017, spanning a period of 34 months. \emph{Gaia} EDR3 data were gathered through the Vizier \citep{Ochsenbein00} service.

This article is organized as follows: In Sect.\,\ref{selection}, we describe the methodology followed to identify candidates to vanishing objects while, in Sect.\,\ref{analysis} we assess the different hypothesis to explain the physical nature of those candidates. After Sect.\,\ref{bd} and Sect.\,\ref{vanishing} where we report the discovery of a brown dwarf candidate in POSS I and give an estimation of the number of vanishing stars, respectively, we summarized the main results of the paper in Sect.\,\ref{Summary}. A brief description of the Virtual Observatory compliant archive than contains detailed information on the candidates is given in the Appendix.

\section{Candidate selection}
\label{selection}

We built a VO-workflow consisting in the following steps: 
\begin{itemize}
    \item Sky tessellation: To  avoid  memory overflow  problems  associated to the data processing of large volumes of data, we made use of the scripting capabilities of Aladin\footnote{\url{https://aladin.u-strasbg.fr/}} \citep{Bonnarel00, Boch14}, to tessellate the sky covered by POSS I in circular regions of 30\,arcmin radius. 
    \item Source extraction: For each one of these regions, we run SExtractor \citep{Bertin96} to build a catalogue of sources. To minimize the number of false sources in noisy images and detect only sources well above the noise level, the {\tt DETECT$_{-}$THRESH} parameter was set to 5 and only sources with SNR$_{-}$WIN>30 and FLAG= 0 were kept. 
    \item Cross-matching: We made use of the CDSSkymatch functionality implemented in STILTS\footnote{\url{http://www.star.bris.ac.uk/~mbt/stilts/}} \citep{Taylor06} to cross-match the catalogue of sources obtained in the previous step with the \emph{Gaia} EDR3 and Pan-STARRS DR2 catalogues. Sources  in the SExtractor catalogue not having counterparts either in \emph{Gaia} or Pan-STARRS in a 5 arcsec radius were kept. The adopted radius is a good compromise to ensure that not many high proper motion were left out while, at the same time, avoiding  an  unmanageable number  of  false  positives.  
    \item Spikes' removal: Spurious detections appearing on the diffraction spikes of very bright sources in POSS I images must be identified and removed. After some trial and error, we implemented the following empirical procedure to remove them:
    \begin{itemize}
        \item For each SExtractor source we look for counterparts in the USNO B-1.0\footnote{\url{https://cdsarc.cds.unistra.fr/viz-bin/cat/I/284}} \citep{Monet03} in a circular region of 90\,arcmin radius. 
        \item SExtractor sources having a USNO counterpart fulfilling any of the following two conditions were rejected. 
        \begin{itemize}
            \item Rmag1 or Rmag2 $<$ 12.4
            \item Rmag1 or Rmag2 $<$ -0.09 $\times$ d + 15.3
        \end{itemize}
        where {\it d} is the angular separation in arcsec between the SExtractor and the USNO B-1.0 source and {\it Rmag1} and {\it Rmag2} are the USNO magnitudes in the red band at two different epochs. (Fig.~\ref{fig:spikes}).
    \end{itemize}
    \begin{figure}
	\includegraphics[width=\columnwidth]{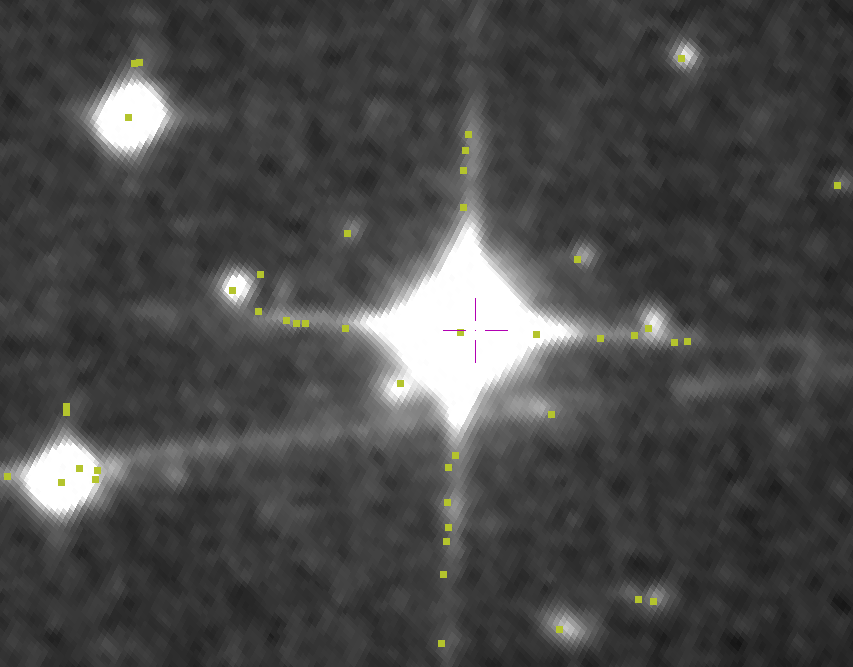}
    \caption{Example of a POSS I image with a bright source showing diffraction spikes. Green squares correspond to USNO B-1.0 sources.}
    \label{fig:spikes}
\end{figure}
\item Other artefacts' removal: Morphometric parameters like the full width at half maximum (FWHM) or the elongation were used to clean the SEXtractor catalogue from spurious sources. For each image, we computed the median and the absolute deviation of the median of the FWHM and the elongation of the sources in order to remove sources deviating more than 2$\sigma$ from the median values (Fig.~\ref{fig:arti}). Additionally, we also applied the following filtering conditions: 
\begin{itemize}
    \item SPREAD$_{-}$MODEL $>$ -0.002
    \item 2 $<$ FHWM $<$ 7 
    \item ELONGATION<1.3
    \item abs((XMAX$_{-}$IMAGE - XMIN$_{-}$IMAGE) -(YMAX$_{-}$IMAGE - YMIN$_{-}$IMAGE)) $<$ 2
    \item XMAX$_{-}$IMAGE - XMIN$_{-}$IMAGE $>$ 1
    \item YMAX$_{-}$IMAGE - YMIN$_{-}$IMAGE $>$ 1
\end{itemize}

{\it SPREAD$_{-}$MODEL} is a SExtractor parameter intended to be a point/extended source classifier. By construction, {\it SPREAD$_{-}$MODEL} is close to zero for point sources, positive for extended sources (galaxies), and negative for detections smaller than the PSF, such as cosmic ray hits. {\it XMAX/XMIN/YMAX/YMIN} indicate the maximum/minimum x/y coordinate among detected pixels. Our condition forces the source to be larger than one pixel and of similar size in both directions. \\

   \begin{figure}
	\includegraphics[width=\columnwidth]{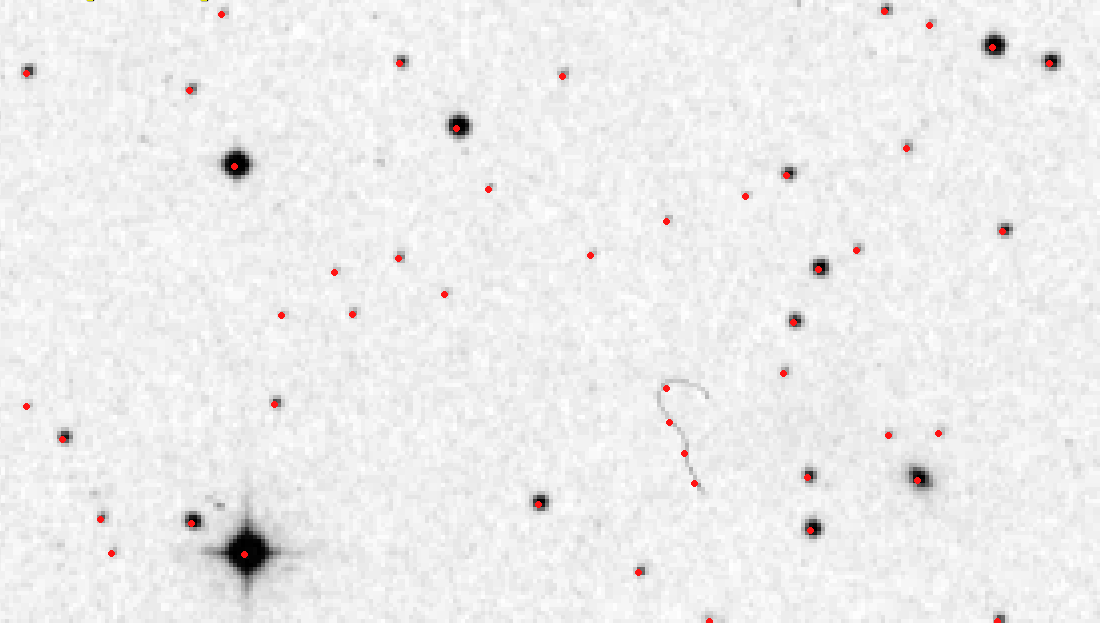}
    \caption{Example of a POSS I image showing a cane-shaped artefact (located at the the centre-right of the image), which is removed from the SExtractor catalogue according to its anomalous values of FWHM and elongation. SExtractor sources are overplotted as red dots.}
    \label{fig:arti}
\end{figure}

\item Identification of high proper motion objects: The $\sim$ 60 yr time baseline between the POSS I images and \emph{Gaia} and Pan-STARRS leads that objects with proper motions typically higher than 80 mas\,yr$^{-1}$ may lie outside our 5 arcmin radius searches becoming, thus, potential candidates. To identify and remove them from the list of candidates, we made use of the observing epoch of the POSS I images (keyword: {\tt EPOCH}). 

For each one of the SExtractor sources fulfilling all the conditions described in the previous steps, we cross-matched them with \emph{Gaia} EDR3 in a 180 arcmin radius, keeping all the counterparts in that radius. For these \emph{Gaia} counterparts, we kept those having proper motion information, which was used to correct the position of the \emph{Gaia} counterparts to the POSS I epoch. The adopted epoch for \emph{Gaia} was J2016.0. SExtractor sources having a \emph{Gaia} counterpart (corrected at POSS I epoch) at less of 5 arcsec were flagged as high proper motion sources and, therefore, removed from the list of candidates (Fig.~\ref{fig:hpm}). \\

   \begin{figure*}
	\includegraphics[width=\textwidth]{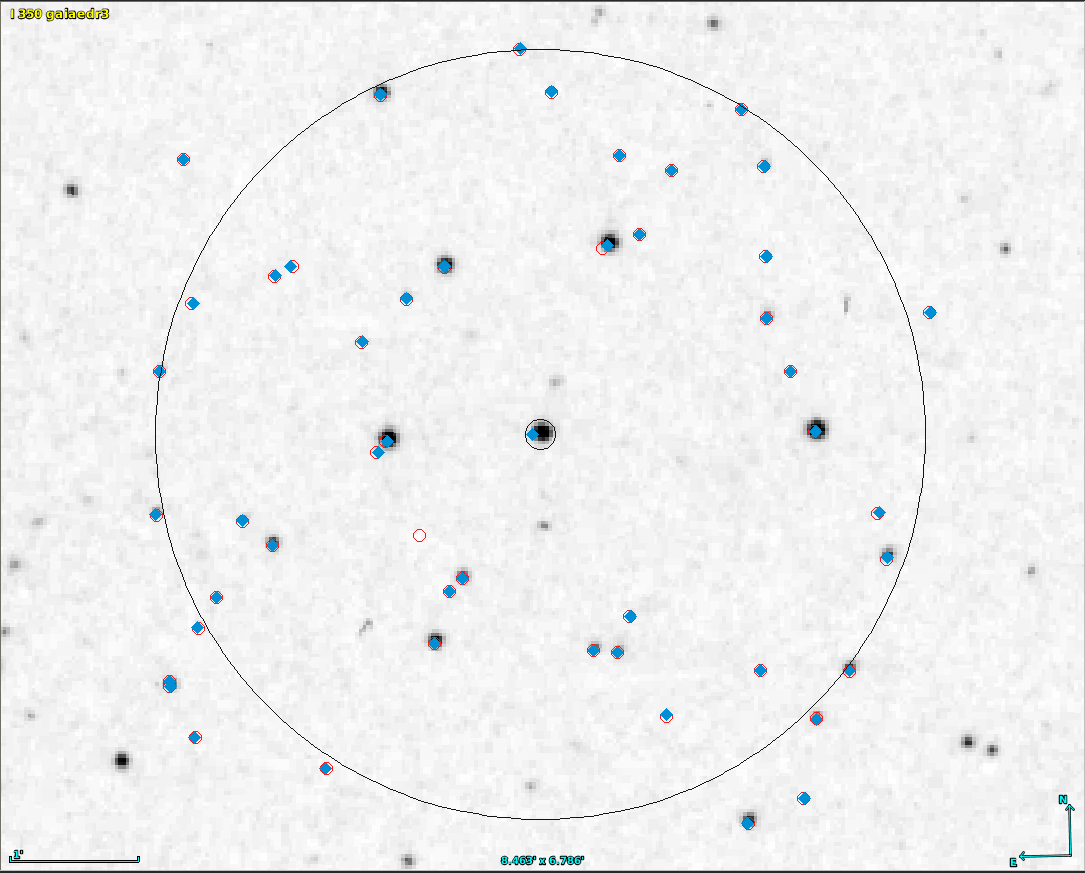}
    \caption{Example of a high proper motion source (centre of the image). Red open circles represent Gaia counterparts at J2016.0 epoch. Solid blue diamonds indicate the position of the same sources at the POSS I epoch. The isolated red circle at lower left of centre really coincides with the POSS I source at the centre of the image if the POSS I epoch is considered (blue diamond inside the inner circle). The outer and inner black circles correspond to the 3 arcmin and 5 arcsec search radius, respectively.}
    \label{fig:hpm}
\end{figure*}

    \item Concatenation: The lists of sources fulfilling all the previous steps were concatenated into a single table and removed duplicated instances as tessellated images may overlap. After all this process, we ended up with a list of 298\,165 POSS I sources not seen either in \emph{Gaia} EDR3 or Pan-STARRS DR2.
    

    
\end{itemize}


\section{Analysis}
\label{analysis}
In the previous section we obtained  a total of 298\,165 potential sources that do not have counterparts either in Pan-STARRS DR2 or \emph{Gaia} EDR3 within a certain distance (5\,arcsec). In what follows we will study the possible nature of these sources. 
\begin{itemize}

\item Objects not present either in \emph{Gaia} EDR3 or Pan-STARRS but present in other astronomical surveys. To check this possibility we looked for counterparts in the catalogues available from the \emph{CDS Upload X-match} utility implemented in TOPCAT \citep{Taylor05} and from IRSA (Neowise, PTF). These searches were complemented with a search in the catalogues available from VOSA \citep{Bayo08}\footnote{The list of catalogues consulted by VOSA can be found at \url{http://svo2.cab.inta-csic.es/theory/vosa/help/star/credits/}}. Sources with counterparts at less than 5 arcsec in any of the queried photometric catalogues were removed from our list of candidates to vanishing objects. 
       
These searches significantly reduced the number of candidates (from 298\,165 to 9\,395). A significant number ($\sim$ 59 per cent) of the identified sources were visible in infrared catalogues (Neowise, \mbox{CatWISE2020}, unWISE, and the infrared catalogues included in VOSA) but not in the optical (KIDS, Skymapper, and the optical catalogues included in VOSA) or the ultraviolet (GALEX). The sources detected in the infrared but not in the optical/ultraviolet are available from the online archive (see Appendix). 

We also queried the Astrographic Catalogue \citep{Urban98} to look for observations previous to POSS I epoch finding no results.\\
             


 
    \item Asteroids: The typical exposure time of the POSS I red images (45-60\,min) and the mean daily motion of Main Belt asteroids (0.1-0.3\,deg d$^{-1}$, larger for nearer asteroids) make that most of these objects leave a stripe on the POSS I red images (Fig.~\ref{fig:ast}). Although due to its elongated shape, most of them are discarded in the search procedure, very slow asteroids mimicking point-like sources could escape from the filtering criteria. In order to also discard these sources we made use of the SkyBoT\footnote{\url{https://vo.imcce.fr/webservices/skybot/}} VO service \citep{Berthier06} to look for known asteroids lying  in the POSS I field of view at the time when the image was observed. Sources lying at less than 1 arcmin from the  predicted position of the asteroid were removed. 189 asteroids were found leaving us with 9\,206 (9\,395-189) candidates. \\ 
   
    \begin{figure*}
		\includegraphics[width=\textwidth]{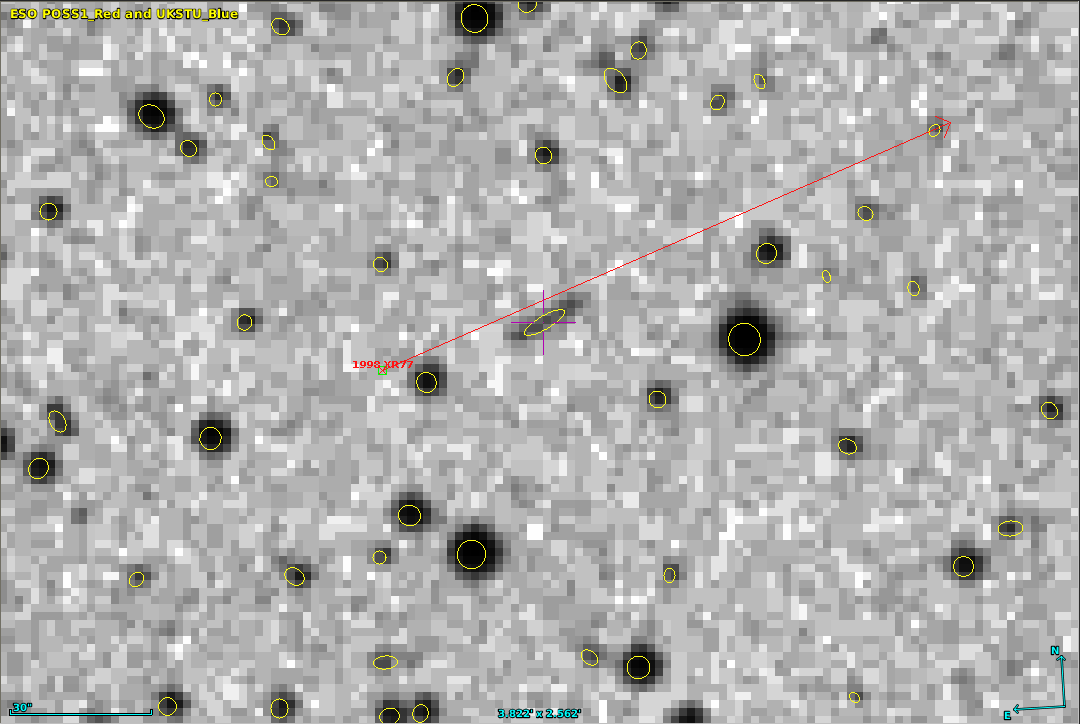}
    \caption{Detection of the asteroid 1998 XR77 (centre of the image). Its elongated, striped shape is clearly visible. The predicted position by Skybot ($\sim$ 35\,arcsec south-est from the real position) as well as its trajectory are plotted in red. Yellow shapes indicate the sources identified by SExtractor.} 
    \label{fig:ast}
\end{figure*}   
 
    \item Stellar variability: 
    It might happen that our candidates are not visible either in \emph{Gaia} EDR3, Pan-STARRS or other catalogues simply because they are large amplitude variable objects reaching their faintest magnitudes at the time when they were observed in those surveys. \emph{Gaia} EDR3 and Pan-STARRS are deeper surveys than POSS I but, how much deeper are they? In order to estimate the limiting magnitude of our POSS I candidates in the \emph{Gaia} EDR3 and Pan-STARRS photometric systems, we randomly selected 100 candidates. Centred on each one of them, we cutout a POSS I image of 15x15 arcmin radius and, using SeXtractor, compiled the sources present in them. A total of $\sim$ 22\,000 sources were obtained. These sources were crossmatched with \emph{Gaia} EDR3 and Pan-STARRS with a 5\, arcsec search radius to look for counterparts in these catalogues. We got 19\,804 and 20\,784 counterparts in \emph{Gaia} EDR3 and Pan-STARRS, respectively. Fig.~\ref{fig:limmag} shows the distribution of the \emph{Gaia} EDR3 ($G$ band) and Pan-STARRS ($r$ band) magnitudes. We see how completeness is reached for POSS I sources at $G$= 18.5 in \emph{Gaia} EDR3 and $r$=19 in Pan-STARRS . The limiting magnitude of \emph{Gaia} EDR3 ($G$ = 21) \citep{Gaia21} and Pan-STARRS ($r$= 21.8)\footnote{\url{https://panstarrs.stsci.edu/}} 
    imply that, for a source not to be visible in \emph{Gaia} EDR3 and Pan-STARRS, the drop in magnitude should be larger than 2.5\,mag approximately. 
    
    Flare stars are objects with spectral types later than late-K that can undergo unpredictable dramatic increases in brightness for a few minutes or hours before returning to their quiescent state  \citep[e.g.][]{Greiner95}. The relatively little time spent at the brightest magnitude (the flare duty cycle for M dwarfs is found to increase from 0.02\% for early M dwarfs
to 3\% for late M dwarfs \citep{Hilton10}. The duty cycle is defined as the percentage of time in a entire period of observation where flares occur), their unpredictable occurrence together with the fact that M stars are the most numerous objects in the galaxy  \citep{Cifuentes2020}, make flare stars good candidates to explain the nature of, at least, a significant part of our sources. Nevertheless, differences in the type of data used to estimate the frequency of flares -- time-resolved photometric \citep{Kowalski09} or spectroscopic \citep{Hilton10} surveys --,  in the wavelength range (near UV, optical), in the magnitudes used to count the number of flares (duty cycle, flares per hour), in the parameters used for flare identification -- {\it flare variability index} \citep{Kowalski09}, {\it flare line index}  \citep{Hilton10} -- , or the existing correlation with spatial distribution, spectral type or age -- begin more frequent close to the Galactic plane, at later spectral types and at younger ages \citep{Hilton10} --, makes it quite difficult to give a realistic estimation of the number of flaring objects among our final list of unidentified transients. 

Other (although less numerous) types of stellar objects triggering large amplitude variations are, for instance, LBVs \citep{Genderen01}, FUORs \citep{Hartmann96}, RCB stars \citep{Benson94}, ILRTs \citep{Cai21}, K giants \citep{Tang10}, cataclysmic objects like novae or pulsating variables like Miras \citep{Reid02}, RV Tau stars \citep{Ruyter05} or Cepheids \citep{Klag09}. Other potential types of variable objects can be found in \citet{Byrne22}. Moreover, extragalactic transients that can be associated to rare blazars or accretion outbursts in active galactic nuclei \citep{Lawrence16} or highly variable quasars or microlensing events \citep{Nagoshi21} can also contribute. 
    
       
   \begin{figure}
	\includegraphics[width=\columnwidth]{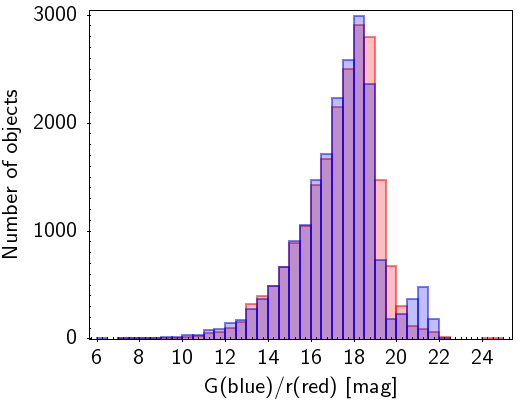}
    \caption{Distribution of the \emph{Gaia} EDR3 (blue) and Pan-STARRS (red) magnitudes of a randomly selected sample of POSS I sources. Limiting magnitudes of POSS I sources at \emph{Gaia} EDR3 and Pan-STARRS are $G$= 18.5 and $r$=19 , respectively.}
    \label{fig:limmag}
\end{figure}

In order to assess whether our candidates to vanishing objects were already known variable objects, we searched for them in the International Variable Star Index\footnote{\url{https://cdsarc.cds.unistra.fr/viz-bin/cat/B/vsx}} \citep{Watson06} and in the Transient Name Server\footnote{\url{https://www.wis-tns.org/}}. In both cases no counterparts were found at less than 5\,arcsec. We also queried the PTF object catalogue available at IRSA\footnote{\url{https://irsa.ipac.caltech.edu/cgi-bin/Gator/nph-scan?mission=irsa&submit=Select&projshort=PTF}} finding 62 counterparts, out of which 35 were flagged as good (\emph{ngoodobs} $>$ 0). This reduced the number of candidates to 9\,171 (9\,206-35).\\
  
  \item Supernovae, hypernovae and failed supernovae: Much less frequent but still possible is that some of our candidates may be ascribed to explosions of very luminous supernovae and hypernovae. If these objects are hosted in very faint galaxies, they will be visible just at the time of explosion keeping no track in images taken subsequently. ASASSN-15lh / SN 2015L, the most luminous supernovae observed so far thought to be originated by the tidal disruption of a star when crossing the tidal radius of a supermassive black hole \citep{Leloudas16}, or GRB171205A a hypernovae caused by the explosion of a very massive star \citep{Izzo19} could be examples of the objects belonging to this category.
  
 Failed supernovae, on its part, have been proposed to explain the absence of Type IIP core-collapse supernovae arising from progenitors above 17\,M$_{\sun}$. In this scenario, the stellar cores will collapse directly to form a black hole without producing the explosion of the star \citep{Byrne22}. As for supernovae and hypernovae, failed supernovae occurring at POSS-I epoch will not be visible in modern archives. Different candidates to failed supernovae have been proposed in the literature \citep{Kochanek08, Reynolds15, Neus21} but the real nature of these objects is still far from being confirmed. \\
  
\item Artefacts: Could artefacts resemble point-like sources and, thus, escape from our morphometric criteria? Small dust particles sticking to the plate during exposures or microspots originated over the years may produce this type of point-like features \citep{Villarroel20}. Clearly, a human inspection of the POSS I plates could solve the problem. However, the original plates of old sky surveys are treated like gold and accessing those plates is extraordinarily rare. An alternative to remove candidates artefacts originated during the scanning process is to compare the images digitized at STSCI with those digitized by SuperCosmos\footnote{\url{https://www.roe.ac.uk/ifa/wfau/cosmos/scosmos.html}}. The comparison was carried out using the Table Access Protocol (TAP) service implemented in the GAVO Data Centre\footnote{\url{http://dc.zah.uni-heidelberg.de/tap}}. Candidates having a counterpart in the Supercosmos digitization at less than 5\,arcsec were kept. After this comparison, 5\,579 candidates still remain. \\

 \item Wrong astrometry. If the POSS I red images from where the candidates to vanishing objects were extracted have poor astrometry, then, the extracted sources will appear displaced from the \emph{Gaia} EDR3 and Pan-STARRS positions being, thus, flagged as candidates. 
    
    To check this possibility, we randomly selected a subsample of $\sim$ 400\,000 sources extracted from POSS I images, sources that were crossmatched with \emph{Gaia} EDR3. The mean and median values for the differences in position were 0.97 and 0.91\,arcsec, respectively with an standard deviation of 0.5\,arcsec. Thus, we can conclude that wrong astrometry cannot explain the number of candidates to vanishing objects we have found. \\
 
    \item Technosignatures: Technosignatures can be defined as properties or effects that cannot be ascribed to natural phenomena and, thus, may indicate an artificial origin (e.g. artificial communication lasers, Dyson spheres and megastructures. In particular, the latter two could make a dim or even vanish entirely the star). The role of vanishing stars in searches for technosignatures were first presented in \citet{Villarroel16}. A general overview describing the possibilities of technosignature searches in time-domain astronomy is given in \citet{Davenport19} while concrete examples can be found in \citet{Villarroel20}. Human satellites at the geostationary orbit could be argued as a possibility to explain the glints found in POSS I images, glints that could be caused by reflections of the Sun. This glints would be bright, have a PSF-like shape and short duration \citep{Villarroel22}. However, we remind the reader that the launch of the first satellite happened in 1957, when most of the POSS I survey was already completed.
 \\
 
\item High proper motion objects without proper motion information in \emph{Gaia} EDR3. Not all sources included in \emph{Gaia} EDR3 have associated an estimation of their proper motion. If this is the case and the object has a significant proper motion, it could be wrongly flagged as candidate to vanishing object. In order to identify these fast-moving objects, we carried out a visual inspection using Aladin. First, we made use of the proper motion information available in Simbad for objects at small angular distances from our candidates. This way we discarded 178 objects (Fig.~\ref{fig:hpmnotgaia}). For the remaining 5\,401 (5\,579-178) candidates we looked for sources at close angular distances in catalogues with different time coverage (2MASS, SDSS, UKIDSS, ZTF) aiming at finding a clear linear displacement between images (Fig.~\ref{fig:hpm1}). After this visual inspection we ended up with a final list of 5\,399 candidates to vanishing objects. A flowchart summarising the selection and analysis process is shown in Fig.\,\ref{fig:workflow}. The spatial distribution in galactic coordinates of the final list of candidates as well the distribution of their magnitudes ($R$ Supercosmos) are given in Fig.\,\ref{fig:gal} and Fig.\,\ref{fig:scdist}, respectively.

 \begin{figure*}
	\includegraphics[width=\textwidth]{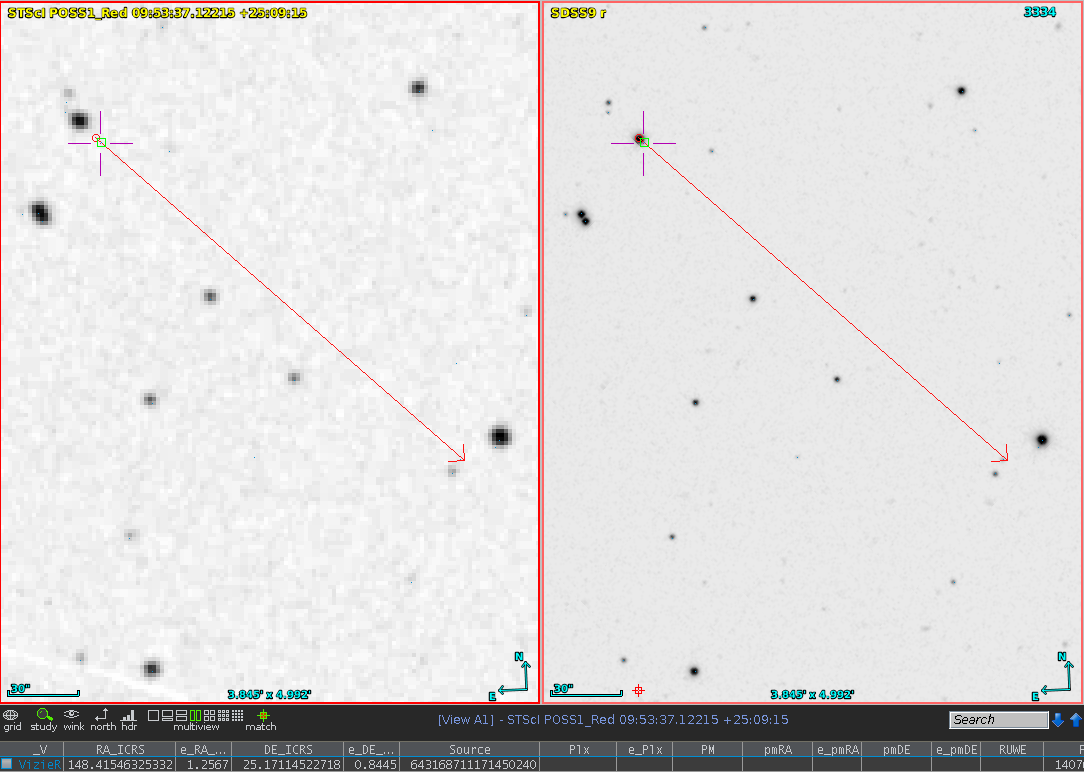}
    \caption{LP 371-1, the object marked with a cross in the SDSS image, is a high proper motion star as reported in Simbad (PMRA: -154\,mas yr$^{-1}$; PMDEC: -140\,mas yr$^{-1}$; red line) but without proper motion information in \emph{Gaia} EDR3 (table shown at the bottom). See text for more details.}
    \label{fig:hpmnotgaia}
\end{figure*}

 \begin{figure*}
	\includegraphics[width=\textwidth]{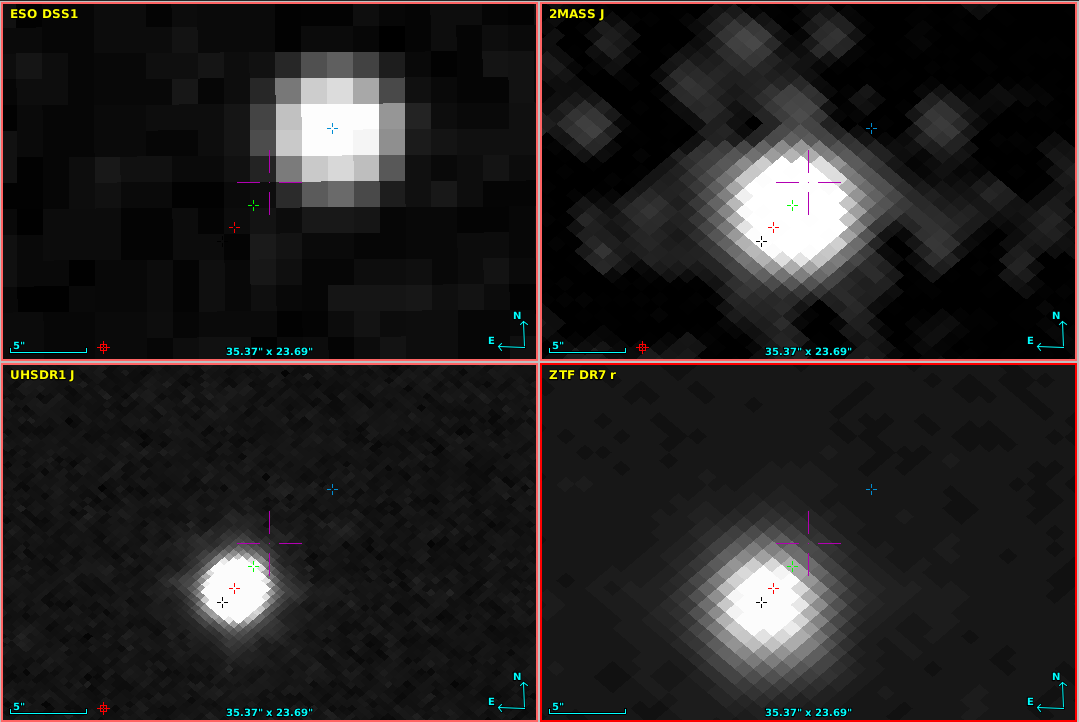}
    \caption{Example of a high proper motion object not reported either in \emph{Gaia} EDR3 or Simbad. The blue/green/red/black crosses mark the position of the source in POSS I/2MASS/UKIDSS/ZTF images, respectively.}
    \label{fig:hpm1}
\end{figure*}

   \begin{figure}
	\includegraphics[width=\columnwidth]{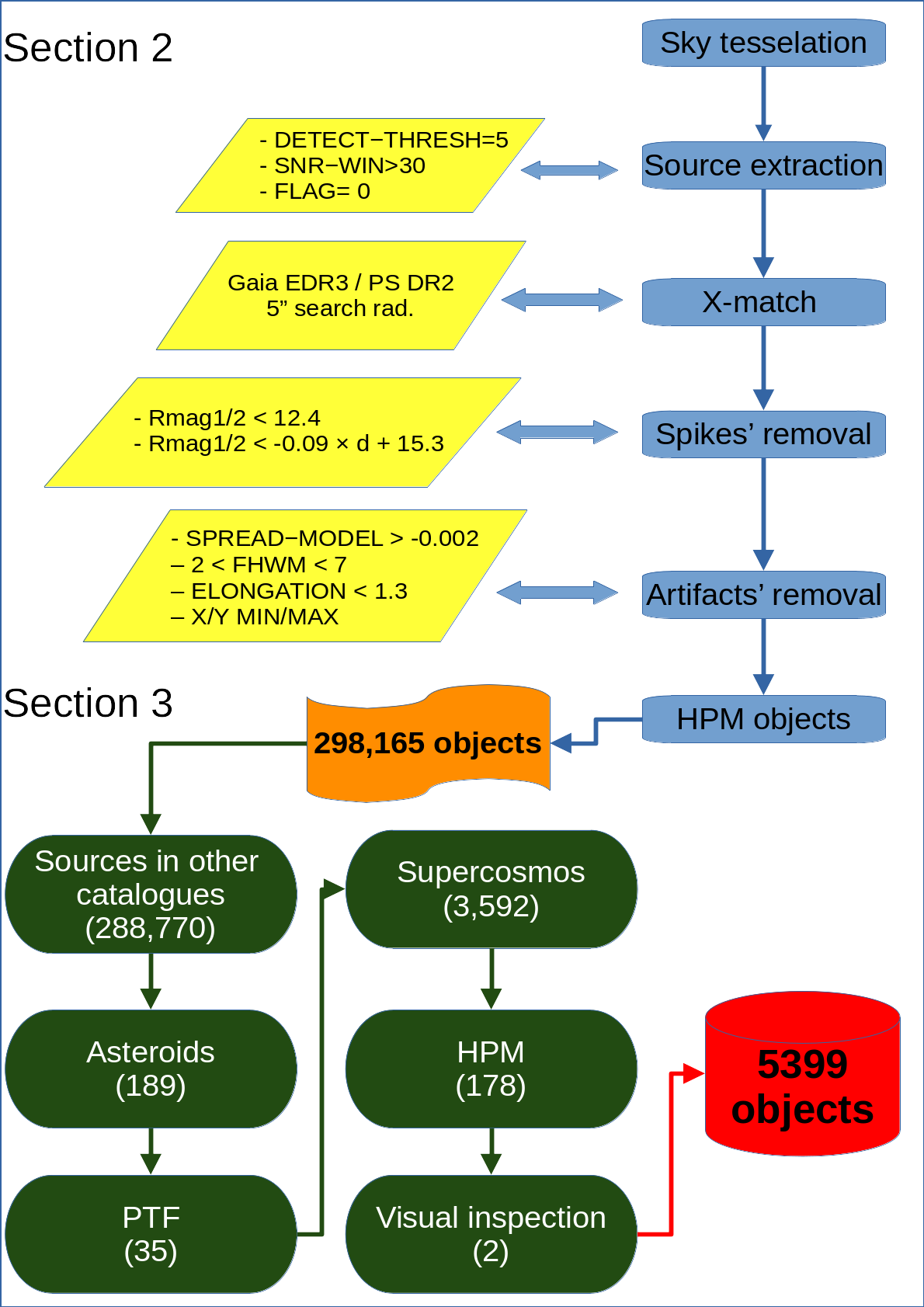}
    \caption{Flowchart of the candidate selection and analysis. See Sect.\,\ref{selection}, and  Sect.\,\ref{analysis} for details.}
    \label{fig:workflow}
\end{figure}

  \begin{figure*}
	\includegraphics[width=\textwidth]{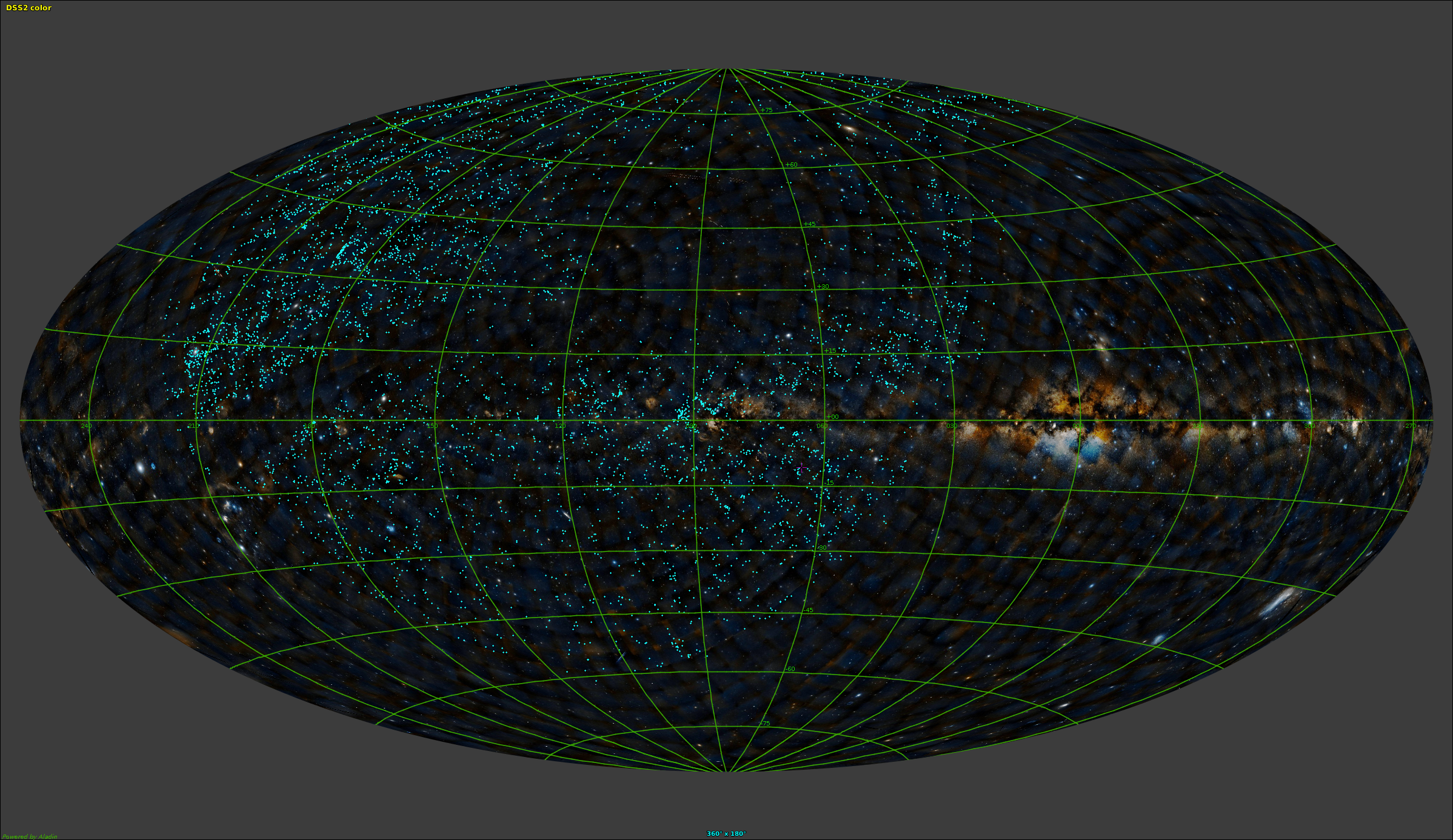}
    \caption{Spatial distribution in galactic coordinates of the final list of 5\,399 candidates (blue dots). A coloured POSS-II images is displayed in the background.}
    \label{fig:gal}
\end{figure*}

  \begin{figure}
	\includegraphics[width=\columnwidth]{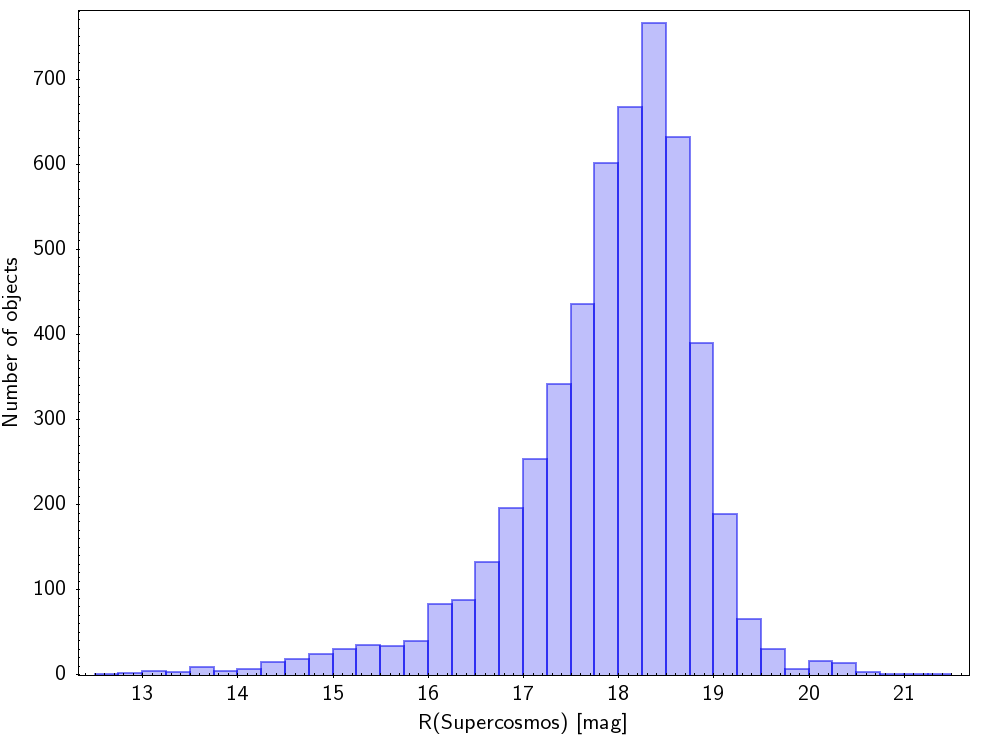}
    \caption{Distribution of the Supercosmos $R$ magnitudes of the final sample (5\,399 objects). It peaks at $R$ $\sim$ 18,5 with 80 \% of the target with magnitudes in the range 17 $\leq$ $R$ $\leq$ 19.}
    \label{fig:scdist}
\end{figure}

\end{itemize}

\section{Brown dwarfs in POSS I images}
\label{bd}
As pointed out in Sect.\,\ref{analysis}, most of the 298\,165 candidates were discarded because they were detected in catalogues other than \emph{Gaia} EDR3 or Pan-STARRS. Since most of the discarded sources were detected in the infrared, we looked into this sample to identify candidate brown dwarfs in POSS I images. This is relevant because it would imply that this class of objects was already recorded in photographic plates forty years before the confirmation of its first member \citep{Rebolo95}. 

\begin{figure*}
    \centering
    \includegraphics[width=10cm]{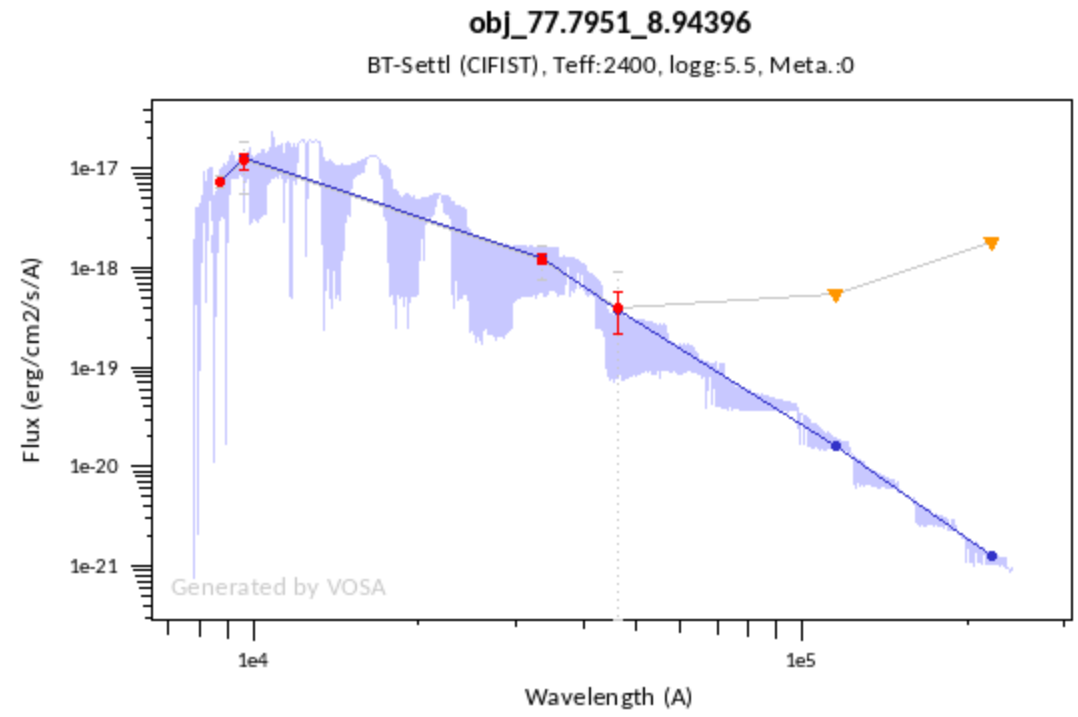}
    \includegraphics[width=\textwidth]{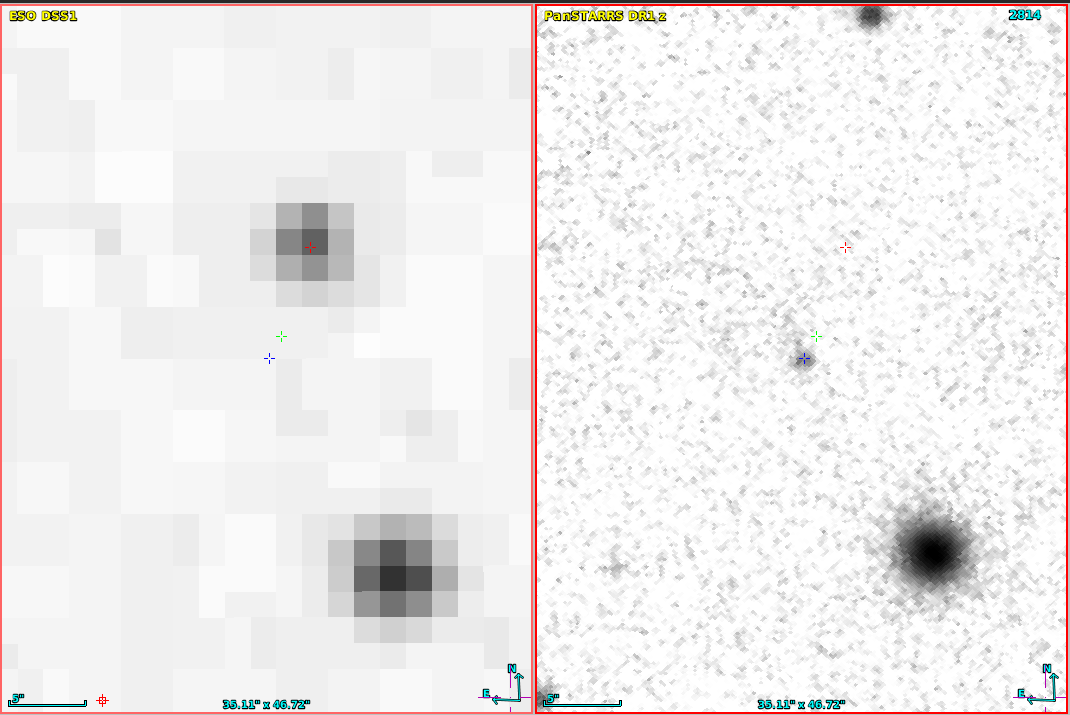}
    \caption{{\bf Top:} Spectral energy distribution fitting of the candidate brown dwarf (RA: 77$^{\circ}$.7951, DEC: 8$^{\circ}$.94396). The red dots represent the observed photometry. Overplotted in blue is the best fitting CIFIST model. Photometric upper limits are plotted as inverted yellow triangles and are not considered in the SED fitting process. {\bf Bottom:} The candidate brown dwarf as seen in POSS I (red cross) and Pan-STARRS (blue cross). The green cross in between indicates the position of the source in an intermediate epoch (2MASS).}
    \label{fig:VOSA}
\end{figure*}

For this, we made a spectral energy distribution fitting using VOSA and the BT-Settl CIFITS grid of model atmospheres \citep{Baraffe15}. A candidate brown dwarf with effective temperature  of 2\,400\,K (M9\,V according to the Mamajek's Color and Effective Temperature Sequence\footnote{\url{https://www.pas.rochester.edu/~emamajek/EEM_dwarf_UBVIJHK_colors_Teff.txt}}) was found. The SED fitting as well the position of the candidate brown dwarf in POSS I, 2MASS and Pan-STARRS images is shown in Fig.~\ref{fig:VOSA}.

\section{Failed Supernovae?}
\label{vanishing}

\begin{figure*}
	\includegraphics[width=14.5cm]{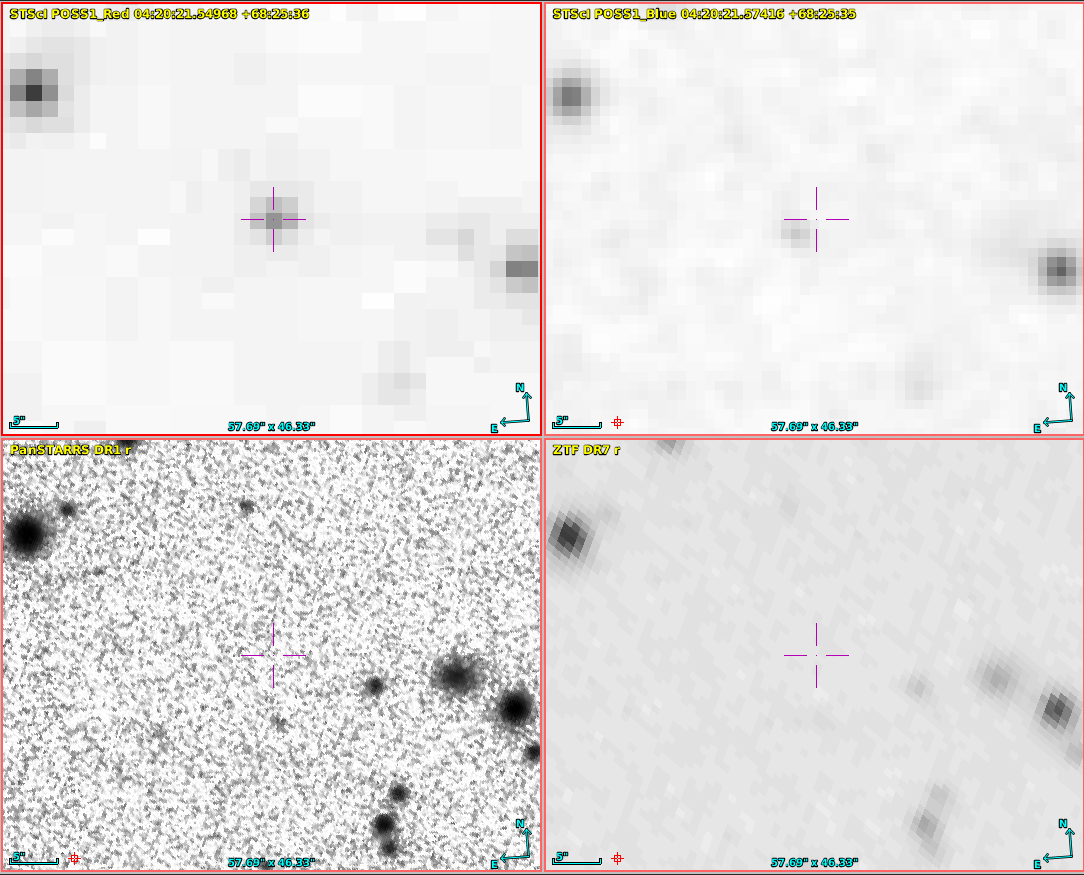}
	\includegraphics[width=14.5cm]{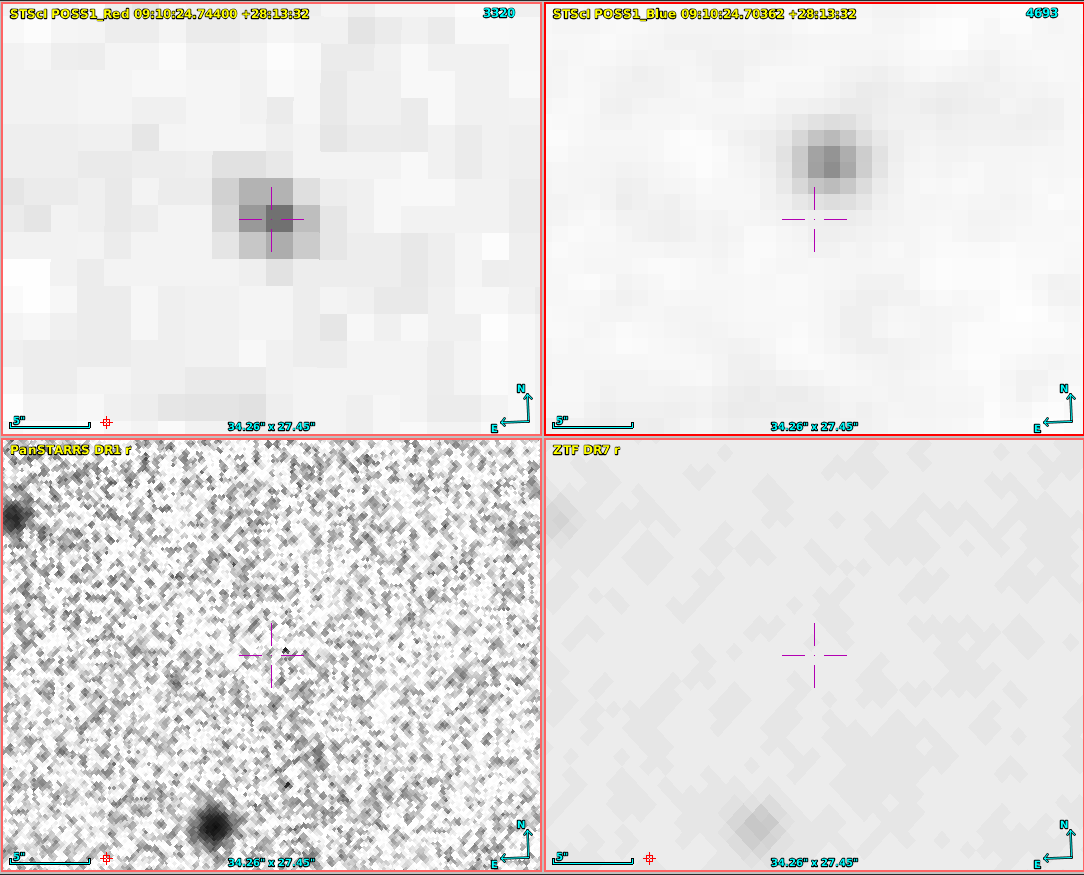}
    \caption{Two sources seen in POSS I red and blue images (differences in time $\sim$ 30\,min) but not detected in more recent surveys.} 
    \label{fig:bluered}
\end{figure*}

In \cite{Villarroel20} the rate of failed supernovae was estimated to be less than 1 in 90 million during a 70 years of time window. We investigate the list of 298\,165 transient sources to see if any of these can be found both in POSS I and in POSS II red images before vanishing.
We find five candidates that each one turns out to be a superposition of a transient and artefact. Also, among the 5\,399 final candidates, we found only two sources almost simultaneously observed in the POSS I blue and red images (the difference in observing time between the blue and red exposures is, typically, $\sim$\,30\,min). Although, on the basis of the displacements of the sources between the POSS I blue and red images, they could be classified as non-catalogued asteroids, the fact that the two sources show a point-like shape might question this hypothesis since, as mentioned in Sect.\,\ref{analysis}, asteroids are expected to show an elongated shape in the direction of the movement (Fig.\,\ref{fig:bluered}). The small number of almost simultaneous transients in blue and red images can be, at least, partly explained by the different wavelength coverage (very red sources may not be visible in the blue, and viceversa) and by the different exposure times (typically, 45-60 min in the red and 8-10 min in the blue) which, despite some flux dilution due to the large observing time,  makes higher the likelihood of finding a transient in red images. 


The above results indicate that an entire disappearance of a star might be rare, 
which agrees with some theoretical predictions \citep{Byrne22}. All-sky survey searches for failed supernovae are more likely to succeed on declining brightness of an object within $\Delta$m$< 3$ mag, rather than its entire disappearance. In order to reach a final conclusion on the rate of failed supernovae, a future study will carefully examine whether the complete list of 298,165 sources, as well as the list of candidates emerging from the VASCO citizen science project, contains candidates visible in both the blue and the red band.

\section{Summary}
\label{Summary}


Working with three large-area sky surveys (POSS I, \emph{Gaia} EDR3 and Pan-STARRS DR2) and a workflow based on Virtual Observatory archives and services, we have searched for sources identified in POSS I but not seen either in \emph{Gaia} or Pan-STARRS finding 298\,165 sources. After filtering sources found in other archives (mainly in the infrared), asteroids, high proper motion objects with no information on proper motion in \emph{Gaia} EDR3, known variables and artefacts, we ended up with a list of 5\,399 sources. Working with POSS I data has the advantage of getting rid from contamination of artificial satellites and, at the same time, opens the possibility of exploring long-term (of the order of decades) variability phenomena.


 Although the origin of these 5\,399 vanishing sources is not clear, most of them might be associated to large amplitude ($>$ 2.5\,mag) variable stars like, for instance, flare stars. Other physical (unknown asteroids, non-catalogued high proper motion objects or exotic objects theoretically proposed like failed supernovae) and artificial (technosignatures) mechanisms can also be proposed to explain the disappearance of our list of objects in modern archives. 
 
 The candidates to vanishing stars 5\,399 as well as the sources detected in the infrared but not in the visible (172\,163) are easily accesible from a Virtual Observatory compliant archive. These sources can be of interest in searches for strong M-dwarf flares, extreme stellar variability on extended times-scales as well as extragalactic transients. Follow-up observations with bigger telescopes will reveal the presence/absence of an object in its place. Sources still missing after follow-up observations could also be useful for technosignature studies like, for instance, laser searches \citep{Villarroel20,Marcy22}. The 298\,165 sources will be further examined by the VASCO citizen science project \citep{Villarroel20b}.
 Finally, it is also important to stress the fundamental role played by the Virtual Observatory in this paper. The discovery, access and analysis of millions of objects coming from tens of archives covering the electromagnetic spectrum from the ultraviolet to the mid-infrared would have not been possible without the tools and services provided by this e-infrastructure.    




\section*{Acknowledgements}

This research has made use of the Spanish Virtual Observatory (https://svo.cab.inta-csic.es) project funded by MCIN/AEI/10.13039/501100011033/ through grant PID2020-112949GB-I00 and MDM-2017-0737 at Centro de Astrobiolog\'{i}a (CSIC-INTA), Unidad de Excelencia Mar\'{i}a de Maeztu. B.V. is funded by the Swedish Research Council (Vetenskapsr\aa det, grant no. 2017-06372) and is also supported by the The L’Or\'{e}al - UNESCO For Women in Science Sweden Prize with support of the Young Academy of Sweden. She is also supported by M\"{a}rta och Erik Holmbergs donation.

This work presents results from the European Space Agency (ESA) space mission \emph{Gaia}. \emph{Gaia} data are being processed by the \emph{Gaia} Data Processing and Analysis Consortium (DPAC). Funding for the DPAC is provided by national institutions, in particular the institutions participating in the Gaia MultiLateral Agreement (MLA). The \emph{Gaia} mission website is https://www.cosmos.esa.int/gaia. The \emph{Gaia} archive website is https://archives.esac.esa.int/gaia. 

The Pan-STARRS1 Surveys (PS1) and the PS1 public science archive have been made possible through contributions by the Institute for Astronomy, the University of Hawaii, the Pan-STARRS Project Office, the Max-Planck Society and its participating institutes, the Max Planck Institute for Astronomy, Heidelberg and the Max Planck Institute for Extraterrestrial Physics, Garching, The Johns Hopkins University, Durham University, the University of Edinburgh, the Queen's University Belfast, the Harvard-Smithsonian Center for Astrophysics, the Las Cumbres Observatory Global Telescope Network Incorporated, the National Central University of Taiwan, the Space Telescope Science Institute, the National Aeronautics and Space Administration under Grant No. NNX08AR22G issued through the Planetary Science Division of the NASA Science Mission Directorate, the National Science Foundation Grant No. AST-1238877, the University of Maryland, Eotvos Lorand University (ELTE), the Los Alamos National Laboratory, and the Gordon and Betty Moore Foundation. \\

This publication makes use of VOSA, developed under the Spanish Virtual Observatory project. This research has made use of \mbox{Aladin} \citep{Bonnarel00, Boch14}, Simbad \citep{Wenger00}, and Vizier \citep{Ochsenbein00}, developed at CDS, Strasbourg Observatory, France. TOPCAT \citep{Taylor05} and STILTS \citep{Taylor06} have also been widely used in this paper.
This research has made use of the NASA/IPAC Infrared Science Archive, which is funded by the National Aeronautics and Space Administration and operated by the California Institute of Technology. This research has also made use of IMCCE's SkyBoT VO tool.

\section{Data availability}
The data underlying this article are available in the SVO archive of vanishing objects in POSS I red images available at http://svocats.cab.inta-csic.es/vanish/




\bibliographystyle{mnras}
\bibliography{example} 




\appendix
\section{Virtual Observatory compliant, online catalogue}
\label{catalogue}

In order to help the astronomical community on using our
catalogue of candidates to vanishing objects,
we developed an archive system  that  can  be  accessed  from  a  webpage\footnote{\url{http://svocats.cab.inta-csic.es/vanish/}} or  through  a
Virtual Observatory ConeSearch.\footnote{for instance, \url{http://svocats.cab.inta-csic.es/vanish-possi/cs.php?RA=0.708&DEC=47.155&SR=0.1&VERB=2}}

The  archive  system  implements  a  very  simple  search
interface that allows queries by coordinates and radius as
well as by other parameters of interest. The user can also select the maximum number of sources (with values from 10 to
unlimited).
The result of the query is a HTML table with all the
sources found in the archive fulfilling the search criteria. The
result can also be downloaded as a VOTable or a CSV  file.
Detailed information on the output  fields can be obtained
placing the mouse over the question mark located close
to the name of the column. The archive also implements the
SAMP\footnote{\url{http://www.ivoa.net/documents/SAMP}}
(Simple  Application  Messaging)  Virtual  Observatory  protocol.  SAMP  allows  Virtual  Observatory  applications  to  communicate  with  each  other  in  a  seamless  and
transparent manner for the user. This way, the results of a
query  can  be  easily  transferred  to  other  VO  applications,
such as, for instance, Topcat.




\bsp	
\label{lastpage}
\end{document}